\UseRawInputEncoding
\documentclass[a4paper,10pt,oneside]{article}
\usepackage[utf8]{inputenc}
\usepackage{icad2023,amsmath,epsfig,times,url,hyperref, amssymb, pifont}
\usepackage[T1]{fontenc}
\usepackage[usenames,dvipsnames]{xcolor}

\definecolor{chgcol}{rgb}{0., 0., 0.}

\newcommand{\strauss}{\texttt{strauss}}
\newcommand{\ttf}[1]{{\texttt{#1}}}
\newcommand{\bttf}[1]{{\textbf{\texttt{#1}}}}
\newcommand{\ittf}[1]{{\textit{\texttt{#1}}}}

\newcommand{\cmark}{\ding{51}}%
\newcommand{\xmark}{\ding{55}}%
\newcommand{\chg}[1]{#1}

\title{Introducing \strauss{}: A flexible sonification Python package}

\twoauthors{\chg{James W. Trayford}} {Institute of Cosmology and Gravitation \\ University of Portsmouth \\ Dennis Sciama Building \\ Burnaby Road \\Portsmouth, United Kingdom  \\ {\tt james.trayford@port.ac.uk}}
{\chg{Chris M. Harrison}} {School of Mathematics, Statistics \& Physics \\ Newcastle University \\ Herschel Building \\ Newcastle upon Tyne, United Kingdom  \\ {\tt christopher.harrison@newcastle.ac.uk}}

\begin{document}
\ninept
\maketitle
\begin{sloppy}
\begin{abstract}
We introduce \strauss{} (\textbf{S}onification \textbf{T}ools and \textbf{R}esources for \textbf{A}nalysis \textbf{U}sing \textbf{S}ound \textbf{S}ynthesis) a modular, self-contained and flexible Python sonification package, operating in a free and open source (FOSS) capacity. \strauss{} is intended to be a flexible tool suitable for both scientific data exploration and analysis as well as for producing sonifications that are suitable for public outreach and artistic contexts. We explain the motivations behind \strauss{}, and how these lead to our design choices. We also describe the basic code structure and concepts. We then present output sonification examples, specifically: (1) multiple representations of univariate data (i.e., single data series) for data exploration; (2) how multi-variate data can be mapped onto sound to help interpret how those data variables are related and; (3) a full spatial audio example for immersive Virtual Reality. We summarise, alluding to some of the future functionality as \strauss{} development accelerates.
\end{abstract}

\section{Introduction}
\begin{figure*}
\centering
          \includegraphics[width=\linewidth]{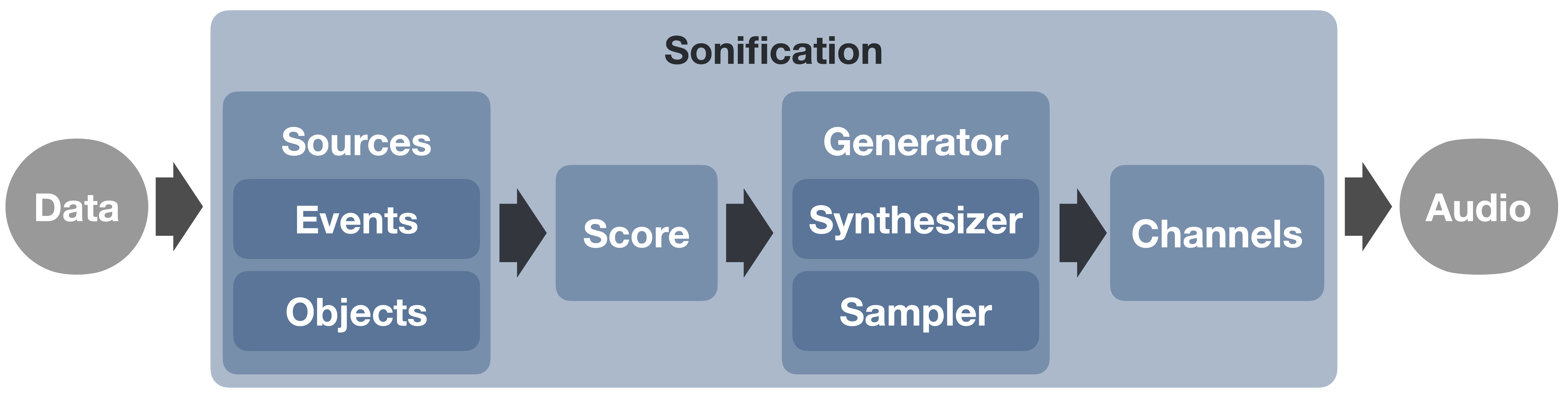}
      \caption{Schematic diagram illustrating the modular workflow of a \strauss{} sonification; going all the way from the raw input data to the output audio representation. The well-defined \textit{sub-modules} can be altered independently, to aid conceptual clarity. The remit of the modules is detailed further in \S~\ref{sec:modules}}.
      \label{fig:flow}
\end{figure*}

\label{sec:intro}

Sonification has been explored for the purposes of data exploration over the many decades of the ICAD community (e.g., see 2012 summary in \cite{Bearman12}). The sonification tools that have emerged for this application come in many guises, including graphical user interfaces (e.g., \textit{StarSound} \cite{Cooke19}, \textit{Psst} \cite{Potluri22}), code-based tools (e.g., \textit{SoniPy} \cite{Worrall07}, \textit{astronify}\footnote{\url{https://astronify.readthedocs.io}}), and recently online hosted tools (e.g., \textit{Highcharts Sonification Studio}, \cite{Cantrell21}, \textit{SonoUno} \cite{DeLaVega23}).

As we discuss in \S~\ref{sec:motivation}, these tools have various goals in terms of intended users and applications. Sometimes \chg{they} are domain specific, i.e., created for applying sonification within a specific scientific domain. The work presented here originated with our goal to apply sonification within the astronomy domain. Over recent years there has been an explosion of interest in developing approaches and tools for sonification in astronomy for research, science communication or education, and for improving accessibility to those who are blind or have low vision; see review in \cite{Zanella2022}. Astronomy is a domain particularly suited to exploring the potential of sonification, due to: (1) the extremely large and often multi-dimensional datasets (e.g. hyper-spectral datacubes); (2) monitoring high-volume data for interesting and `transient' events (e.g. for  black holes colliding or supernovae explosions); (3) exploring time-series data for repeating features and patterns (e.g., planets transiting in front of stars\cite{TuckerBrown2022}) and; (4) the general popularity of astronomy in the general public and media. As an example of the last point, NASA released a sonification of data from the gas around a black hole, which received international media attention and millions of hits on the Guardian's YouTube channel\footnote{\url{https://www.youtube.com/watch?v=_tXhBLg3Wng}} alone. 

Here we introduce the design, structure and functionality of our \chg{new sonification package} \strauss{}\footnote{\textcolor{chgcol}{\url{https://audiouniverse.org/research/strauss}}} (\textbf{S}onification \textbf{T}ools and \textbf{R}esources for \textbf{A}nalysis \textbf{U}sing \textbf{S}ound \textbf{S}ynthesis). Although originally designed with astronomical data and communication in mind, we demonstrate how it has developed into a Python-based tool with utility for a wide range of applications, and is suitable for use across scientific domains. 

\section{STRAUSS Motivation \& Philosophy}
\label{sec:motivation}
To motivate the design choices of the \strauss{} code, it is important to first motivate the needs it is aiming to fulfil, as well as  placing it in the context of pre-existing tools and resources.

We posit that realising the potential of sonification for data-driven fields requires sonification techniques to become embedded into standard practice at the level of the \textit{data analyst} - i.e. those handling and interpreting data. The data methodologies and modalities used by analysts set the perception and standards of their fields. Analysts can also represent a great range of experience levels; from students working on first projects, to world experts conducting groundbreaking research. Furthermore, the data analyst may wish to produce sonification both for their own data exploration and to communicate their research to a wide range of audiences: from the general public or school pupils through to other experts in their field. Having analysts familiar and comfortable with sonification, while thinking creatively about expression of their data through sound (beyond sonification as a passing novelty), may unlock sonification's transformative benefits.

For familiarity and comfort to come about, we aimed to create an approach which makes everyday sonification practical, while flexible enough to handle diverse and creative applications. Furthermore, we aimed for the tool to cater to the broad experience and aptitude range of potential users.

 We use the description \textit{``low barrier, high ceiling"} for such a tool; accessible and understandable for a new user to quickly perform elementary tasks, while allowing detailed technical control for the expert. We set out to create a tool that could be \textit{easily integrated} into the analyst's workflow, i.e. is not disruptive to use. This could include incorporating sonification as an option into the software that analysts already use. This aims to make it as straightforward as possible for analysts to start sonifying data, especially for users who are less familiar with sonification or unaware of its potential. Finally, we aim for a sonification process that is \textit{conceptually clear}, to avoid alienating newcomers curious about sonification. This can be achieved by elements of the tool being well partitioned and structured, with a \textit{coherent design philosophy}. Together, these criteria motivate \strauss{}.

In the landscape of existing tools, we find an \chg{unserviced} niche for \strauss{} between specialised tools for single, specific sonification applications (e.g. pitch mapping data series, inverse spectrograms of images) and highly generalised audio tools (digital audio workstations, audio synthesis environments, etc). The former typically perform one task well but offer limited flexibility (and therefore creative potential), the latter enables much finer control and broader possibilities, but typically requires users to build their own procedures with data-audio interface and mapping, separate from their usual workflow. This can be thought of as a gap between the novice being introduced to sonification and the seasoned expert, creating their own audio experiments. \strauss{} is intended to bridge this gap for the typical analyst. It aims to enable the data analyst to develop from a sonification newcomer to expert, without necessarily breaking out from their normal workflow, and fostering broader experimentation with and popularity of auditory display in data-driven fields. \chg{The general functionality of \strauss{} is designed to extend far beyond pre-existing tools for analysts}. 

We also acknowledge tools that provide their own self-contained environment and interface, also  enabling visually impaired (VI) users \cite{Potluri22, Cantrell21, DeLaVega23,Casado23}. These are crucial for accessibility but fill a different role to \strauss{} - we intend \strauss{} to primarily function to popularise and facilitate sonification widely among analysts using code-based workflows, as a means to improve accessibility. Meanwhile, we do also intend to improve the accessibility of \strauss{} itself (see \S~\ref{sec:usage}) and encourage its incorporation to provide accessible functionality for existing software.

\section{Design choices}
\label{sec:design}

For easy integration and wide applicability, we opted to make \strauss{} into a Python package. This has many benefits. Python is a widely used coding language, boasting a market share of 44\% among software developers and tied for most desired language to learn for existing developers in the \textit{Stack Overflow 2022 User Survey}  \cite{SO22}. Python makes external libraries very easy to incorporate, installed via a package manager (e.g. \texttt{pip}) and incorporated into code using the \texttt{import} system. For less technical users, graphical user interface (GUI) programs written in Python could simply import such a package to add sonification functionality. 

To consider the actual functionality of \strauss{}, a useful analogy to draw is to \textit{visualisation} and specifically to Python \textit{plotting} packages. In Python, this is exemplified by libraries such as \texttt{matplotlib}\cite{Hunter07} and \texttt{yt}\cite{Turk11}. With the broad functionality and low-level control they afford, these libraries can be used by analysts to produce anything from plots prioritising the conveyance of technical information, to images for outreach applications prioritising aesthetics. While we should be cautious of this analogy - visualisation tools do not necessarily map directly to sonification, and sonification may be enhanced by wholly new approaches - this offers a useful template for our design approach, and we make use of this analogy throughout.

Along with this template, the \strauss{} \textit{design philosophy} benefits from a clearly delineated, modular workflow illustrated in Fig~\ref{fig:flow}. An \textit{object-oriented paradigm} (OOP) allows us to represent these modules using Python classes, such that they are entirely self-contained, and can be operated on independently, while collected together in the overarching \texttt{Sonification} class. These constituent parts of \strauss{} are explained fully in \S~\ref{sec:modules}. 

\section{Code Structure \& Functionality}
\label{sec:modules}

Given the general context and motivation of the code, we now provide a brief overview of components of a \strauss{} sonification, represented by modules of the \strauss{} package. A flow diagram of the different modules is shown in Fig.~\ref{fig:flow}. More extensive, evolving documentation for \strauss{} is hosted alongside the code\footnote{\url{https://strauss.readthedocs.io}}. \chg{The online documentation should provide more technical and practical detail on using \strauss{} and its underlying algorithms, relative to the introductory overview provided here.}

\subsection{The \texttt{Sources} module}
\label{sec:sources}

\begin{table}[]
    \centering
    \footnotesize
\begin{tabular}{ l || c | c | c }
\hline
Source parameter & Subject & Map & Evolve \\
\hline 
\chg{\ittf{polar}} & \ttf{Channels} & \cmark & \cmark  \\
\chg{\ittf{azimuth}} & \ttf{Channels} & \cmark & \cmark  \\
\ittf{volume} & \ttf{Sources} & \cmark & \cmark  \\
\ittf{pitch} & \ttf{Sources} & \cmark & \xmark  \\
\ittf{time} & \ttf{Score} & \cmark & \xmark  \\
\ittf{cutoff} & \ttf{Generator} & \cmark & \cmark  \\
\ittf{time\_evo} & \ttf{Score} & \xmark & \cmark  \\
\ittf{spectrum} & \ttf{Generator} & \cmark & \xmark  \\
\ittf{pitch\_shift} & \ttf{Generator} & \cmark & \cmark  \\
\ittf{volume\_envelope/A} & \ttf{Generator} & \cmark & \xmark  \\
\ittf{volume\_envelope/D} & \ttf{Generator} & \cmark & \xmark  \\
\ittf{volume\_envelope/S} & \ttf{Generator} & \cmark & \xmark  \\
\ittf{volume\_envelope/R} & \ttf{Generator} & \cmark & \xmark  \\
\ittf{volume\_lfo/freq} & \ttf{Generator} & \cmark & \xmark  \\
\ittf{volume\_lfo/freq\_shift} & \ttf{Generator} & \cmark & \cmark  \\
\ittf{volume\_lfo/amount} & \ttf{Generator} & \cmark & \cmark  \\
\ittf{pitch\_lfo/freq} & \ttf{Generator} & \cmark & \xmark  \\
\ittf{pitch\_lfo/freq\_shift} & \ttf{Generator} & \cmark & \cmark  \\
\ittf{pitch\_lfo/amount} & \ttf{Generator} & \cmark & \cmark  \\
\hline
\end{tabular}
\caption{Table of \ttf{Sources} parameters currently available for data mapping in \strauss{}, and the  subject class. \textit{`Mappable'} properties can be mapped to persistent values per \ttf{Sources}. \textit{`Evolvable'} parameters can also be parameterised to a data series, for continuous evolution via the \ttf{Objects} source type.}
    \label{tab:mapping}
\end{table}

In \strauss{}, data is represented through instances of the \texttt{Sources} class.
Particularly, \texttt{Sources} specify the mapping between numerical values of variables in the data and the expressive properties of sound in the output sonification. In the \strauss{} paradigm, sound emanates from these \texttt{Sources}, allowing us to naturally represent spatial data in conjunction with the \texttt{Channels} class (\S~\ref{sec:channels}). A \strauss{} sonification can comprise multiple sources, with each source able to support many variable mappings from the data. This provides a flexible system for inputting data into \strauss{}, with \texttt{Sources} parameters used internally throughout the code when generating the sonification. In \strauss{}, the user can choose to assign their data to two different types of sources class, \texttt{Events} and \texttt{Objects}, detailed below. Example applications of each type are demonstrated in \S~\ref{sec:examples}.

Some of the key mappable properties are the direction to the source with respect to the \textit{listener} (specified via  \chg{\texttt{azimuth}} and \chg{\texttt{polar}} angles) the base \ttf{pitch} of the source, the relative \ttf{volume} of the source and the relative \ttf{time} in the sonification that the source is triggered. A full list of these mappable parameters can be found in Table~\ref{tab:mapping}. The correspondence between the mapped data  and source parameter value ranges can also be specified when setting up the \texttt{Source}, either as a percentile (0-100\%) or in data units. By default, \strauss{} maps the full range of the mapped data (from 0 to 100\%) to the full \ttf{Sources} parameter range. 

\textbf{\texttt{Events}:} The \texttt{Events} child class is used to represent sources defined by some \textit{occurrence \ttf{time}} and singular event properties. These can be thought of as representing individual data \textit{points}. Some examples of data well-fit to the \texttt{Events} class could be, supernovae exploding, lightning strikes or mutations in a genome. Table~\ref{tab:mapping} lists the currently implemented `mappable' properties for \ttf{Events} sources, which can be parameterised once per source (e.g. mapping a single \ttf{pitch} to the maximum brightness of a set of supernova events).

\textbf{\texttt{Objects}:} By contrast to \texttt{Events}, \texttt{Objects} \textit{persist through time}, with properties that \textit{evolve}. These can be thought of as as a means to represent data \textit{series} with a single source. Some example data fitting this description could be a galaxy evolving through cosmic time, a hurricane developing in the Atlantic or a stock price changing over a year. \texttt{Objects} support the same `mappable' properties specified once per source in Table~\ref{tab:mapping}, but also a subset of `evolvable' sources parameters that can be evolved throughout the sonification. 

\subsection{The \texttt{Score} module}
\label{sec:score}
With the audio \texttt{Sources} defined, the \texttt{Score} class allows us to place `musical' constraints on the sound they produce to represent the underlying data. The duration of the output sonification is also specified via the \ttf{Score} with the timeline of the the sonification scaled to fit this duration. \strauss{} primarily supports a \textit{`chordal'} score, where a chord or sequence of chords (i.e. notes to be played simultaneously) can be specified. The `\ittf{pitch}' \ttf{Sources} property mapping (Table~\ref{tab:mapping}) is then used to assign \ttf{Sources} to notes, such that each source articulates (or \textit{`arpeggiates'}) a note in the chord. By default this is achieved by `binning' the mapped pitch parameters onto each note in ascending order, such that an approximately equal number of sources represent each note in the chord\footnote{i.e. for a $n$-note chord, defining bins using $n+1$ evenly spaced \textit{percentiles} of the \ttf{pitch}-mapped data variable.}.Where sequences of chords are specified, the chords change through the sonification, such that the note articulated \chg{by each of the \ttf{Sources}} also depends on the mapped \ittf{time} parameter. 

Single note sequences can be parameterised as a special case, using chords with a single note or series of single notes. More \textit{`freeform'} or \textit{atonal} sonifications can also be created by using the `\ttf{pitch\_shift}' \chg{\ttf{Sources}} property applied on top of the `\ittf{pitch}' property, allowing the sound frequency of each source to instead be freely varied from a base note over a specified range.

The \chg{\ttf{Sources}} class is relatively underdeveloped in the current incarnation of the \strauss{} code, with plans to add many more options and functions. For example, chord or key changes could be another avenue to express \textit{metadata}, such as epochal changes in datasets. The \chg{\ttf{Sources}} class could also be a place to extend our presetting approach (described in \S\ref{sec:generator}) to improve \strauss{} as a \textit{`low barrier, high ceiling'} tool, such that users without much musical experience could try collections of pre-defined \ttf{Score} configurations, with simple, evocative names, choosing their preference. 

\subsection{The \texttt{Generator} module}
\label{sec:generator}
Knowing how the \texttt{Sources} express the data, and what they will play via the \texttt{Score}, it is now possible to generate the sound. The \texttt{Generator} class takes instruction from the two prior classes and generates audio for each individual source. This can be achieved using either the \texttt{Sampler} or \texttt{Synthesiser} child classes (along with the \ttf{Spectraliser} special case), detailed below.

\bttf{Sampler:} This class generates audio by triggering pre-recorded audio \textit{samples}. A directory of audio files is used to specify which sample to use for each note of the sampler. These samples are loaded into the sampler and are  interpolated to allow arbitrary pitch shifting. Samples can also be looped in a number of ways to allow notes to sustain perpetually. Samples can be use to generate sonification with a more `natural' sound, using familiar instruments such as bells or strings.

\bttf{Synthesiser:} This class instead generates audio additively using mathematical functions via an arbitrary number of oscillators. The \strauss{} synthesizer supports a number of oscillator forms (see preset example below). the \ttf{Synthesizer} can be typically used for less organic but more flexible sonification, without requiring audio samples to be collected or cleared for use. 

\bttf{Spectraliser:} A special case of the \ttf{Synthesizer}, this generator synthesizes sound from an input spectrum, via an inverse Fast Fourier Transform (IFFT), with randomised phases. The user can specify the audible frequency range that the `spectralised' audio is mapped over. 

These generators share a number of modulation options to affect the generated audio. In particular low frequency oscillators (LFO) affecting \textit{volume} and \textit{pitch} can be used to add \textit{`vibrato'} or \textit{`tremolo'} to notes. \strauss{} can also shape the volume \textit{envelope} of a note, via \textit{attack}, \textit{decay}, \textit{sustain} and \textit{release} (ADSR) parameters. Some of these parameters are mappable or evolvable in \strauss{} (Table~\ref{tab:mapping}). These are familiar features in digital audio, with more that could be added to later versions of \strauss{} \cite{Boulanger00}.

There are many potential options that can be varied to specify the sound qualities of the \ttf{Generator} for technical users. For less technical users, \strauss{} uses a preset system, where sets of configuration options can be chosen from by simply specifying their names.  Presets are written in YAML, a common and intuitive configuration file format, allowing nested parameters. In this way, it is envisaged that users and community members could develop their own presets for reuse or contribute them for posterity. An excerpt from the default \ttf{Synthesiser} preset can be seen below:

\begin{scriptsize}
\begin{verbatim}

# preset name
name: "default"

# full description
description: >-
  Default preset for the synthesizer, using three saw 
  wave oscillators, two of which are detuned slightly 
  higher and lower respectively, with lower volumes. This
  gives a harmonically rich sound, suitable for filtering, 
  with detuned unison saws removing some harshness.
# oscillator information
oscillators:
  # oscillator are denoted osc<n> with n=3 by default
  #
  # level: the intrinsic volume
  #
  # detune: the change in tuning as a percentage of the 
  # input frequency
  #
  # form: the waveform, choose from:
  # ['saw', 'square', 'sine', 'tri', 'noise']
  #
  osc1:
    form: 'saw'
    level: 1.
    detune: 0.
    phase: 0
  osc2:
    form: 'saw'
    level: 0.5
    detune: -2.
\end{verbatim}
\begin{center}
    {\Large ...}
\end{center}
\begin{verbatim}
# or 'tremolo'
volume_lfo:
  use: off
  wave: 'sine'
  amount: 0.5
  freq: 3
  freq_shift: 0
  phase: 'random'
  A: 0.
  D: 0.1
  S: 1.
  R: 0.
  Ac: 0.
  Dc: 0.
  Rc: 0.
  level: 1

# Master volume
volume: 1.

# Default pitch selection
pitch: 1.
\end{verbatim}
\end{scriptsize}

The full version of this file is viewable via the \strauss{} repository\footnote{\scriptsize \textcolor{chgcol}{\url{https://github.com/james-trayford/strauss}}}.

\subsection{The \texttt{Channels} module}
\label{sec:channels}
Once sound has been produced for each source, the final step is to mix the audio down into some multi-channel audio format. The \texttt{Channels} class essentially represents a bank of virtual microphones, with 3D antennae patterns, that each correspond to a channel in the output file. This allows us to naturally capture the full 3D sound field from a listening point in the \strauss{} simulation, for physically accurate and realistic spatialisation. \strauss{} allows fine control over how these channels are configured, but it is foreseen that most users will use a pre-configured standard \textit{audio system}, such as  `\texttt{\textit{mono}}', `\texttt{\textit{stereo}}', `\texttt{\textit{5.1}}', `\texttt{\textit{7.1}}', etc. Mixing audio signals between \ttf{Channels} happens on a source-by-source basis, where the direction to \chg{each of the sources} $j$ at a given instant $t$ allows us to compute a coefficient $A_i(t)$ from the antenna pattern of each channel, $j$, which modulates the signal, $S$, added to each channel, $C$, as $C_j(t) = \sum^{n}_{i=0} A_j(t)S_i(t)$. 

These audio systems typically use a \textit{cardioid} antenna, with the exception of the \texttt{\textit{mono}} class, using a single omni-directional (spherical) pattern. A special class of \textit{audio system} provided by \strauss{} is the \textit{ambisonic} systems. In this case, \strauss{} uses a virtual microphone bank with the precise antenna pattern corresponding to spherical harmonics, up to the chosen order, in standard \textit{ambiX} format \cite{Nachbar11}. In this way, \strauss{} can natively provide ambisonic output to arbitrary order e.g. orders 1, 2, 3 yielding 4, 9 and 16 channels respectively. This is demonstrated in \S~\ref{sec:stars}.

\subsection{The \texttt{Sonification} module}
\label{sec:sonification}
The top-level \texttt{Sonification} class loads in all the above classes and produces the final sonification. Once \ttf{Sources}, \ttf{Score}, \ttf{Generator} and \ttf{Channels} classes are defined, the \ttf{Sonification} class is invoked. The \ttf{render()} method can then be run to produce the sonification. Finally, the \texttt{Sonification} class flushes the audio samples from the \texttt{Channels} class to an output audio file, or streamed audio for display in Python notebook format. 

\begin{figure*}
\centering
          \includegraphics[width=\linewidth]{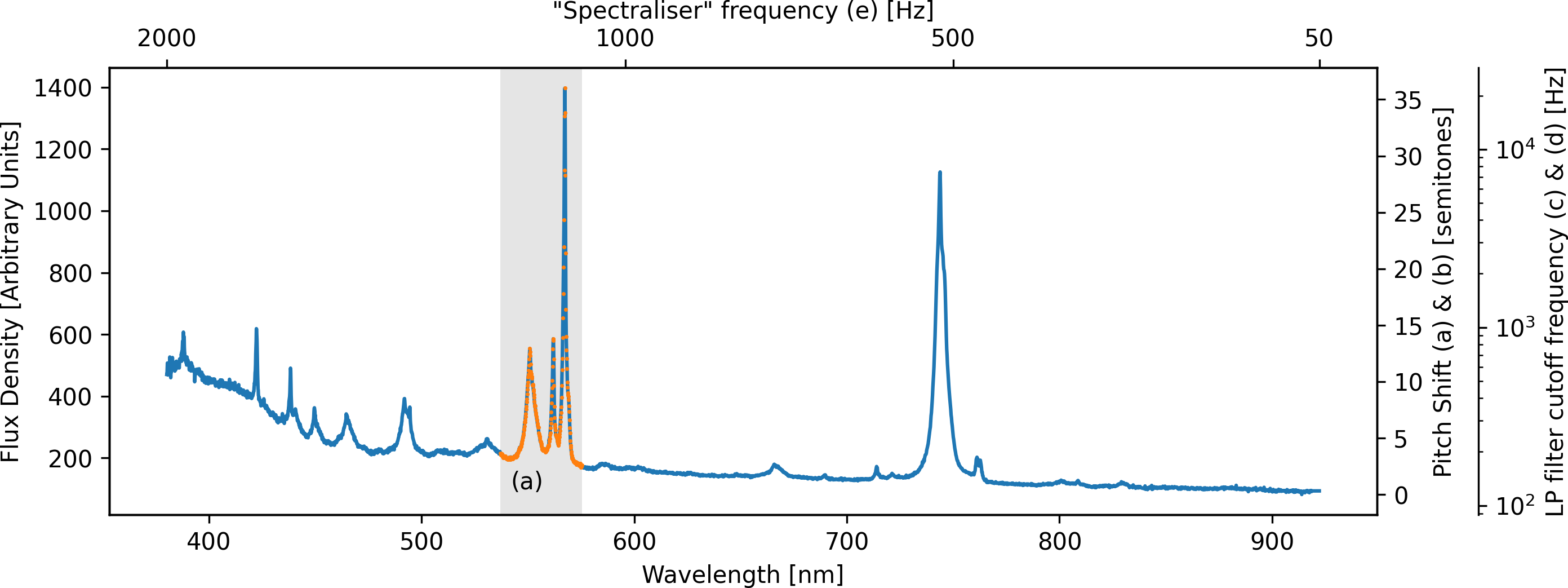}
      \caption{The example astronomical light spectrum of a galaxy (i.e., flux density, or brightness, as a function or wavelength), which has been passed through \strauss{} to illustrate some of the multiple ways of sonifying \textit{univariate data series}. The accompanying axes, to represent the different examples of mapping the data are labelled (a)-(e). The shaded region is the limited wavelength region used for example (a). Each of the methods are expanded upon in Section~\ref{sec:spectrum}, where links are also provided to the video files containing both the audio and a visual representation.}
      \label{fig:spectrum}
\end{figure*}

\section{Using \& Developing \strauss{}}
\label{sec:usage}

We aim to make using \strauss{} as intuitive as possible, with a number of tutorial examples packaged with the code to build this intuition. With \strauss{} targeted to introduce many of its users to its underlying concepts (i.e. those of sonification), clear tutorials are an essential element of the code development. We adopt a \textit{Tutorial Driven Development} (TDD) paradigm; when adding new functionality we first design a model tutorial for its use and \textit{then} implement changes in the code to ensure this works as designed. The reasoning behind this is that our intuition on how the code should work is preserved. Tutorials are available in the \ttf{examples/} subdirectory of the \strauss{} repository, completing these should provide a broad introduction to the practical use of the code.

As we are aiming for a `\textit{low-barrier, high-ceiling}' tool in \strauss{}, it is important to balance technical functionality with intuitive usability. One way we attempt this is via a `presetting' approach, with pre-configured options or the constituent classes of \strauss{}. This is exemplified by the \ttf{Generator} and \ttf{Channels} classes, where simple named presets can be chosen from for less technical users. The preset classes can also be modified to allow full control from a usable starting point. We plan to extend this to the \ttf{Score} and \ttf{Sources} classes, where common musical constraints and mapping choices could be chosen from. This  could be considered analogous to a plotting package like \ttf{matplotlib}, where a \textit{`colorbar'} can be chosen from a predefined set, with evocative names (e.g. `\ittf{winter}', `\ittf{magma}', etc).   

The question of \textit{accessiblity} for users the code is complex. In one sense the \ttf{jupyter} notebook environment aids accessibility, by lowering the barrier to use (allowing \strauss{} and dependencies to be installed in-browser, using cloud resources via e.g. the \textit{Google} \chg{\ttf{Colab}} environment \footnote{\chg{\scriptsize see \url{https://www.audiouniverse.org/research/strauss}}}) and simplifying elements of interface (allowing media to be embedded and played back directly in-notebook, allowing clearly formatted text via markdown).  However, the \ttf{jupyter} environment has a number of issues with accessibility from a visually impaired (VI) standpoint; notebooks are not yet generically accessible with screen readers and other assistive technologies \cite{jupyter-access}. For VI accessibility, future work will go into complementing \ttf{jupyter} tutorials with plain Python script counterparts (which are generically accessible), as well as adapting to the evolving accessibility status of \ttf{jupyter}.

\section{Example applications}
\label{sec:examples}

\begin{figure*}
\centering
          \includegraphics[width=\linewidth]{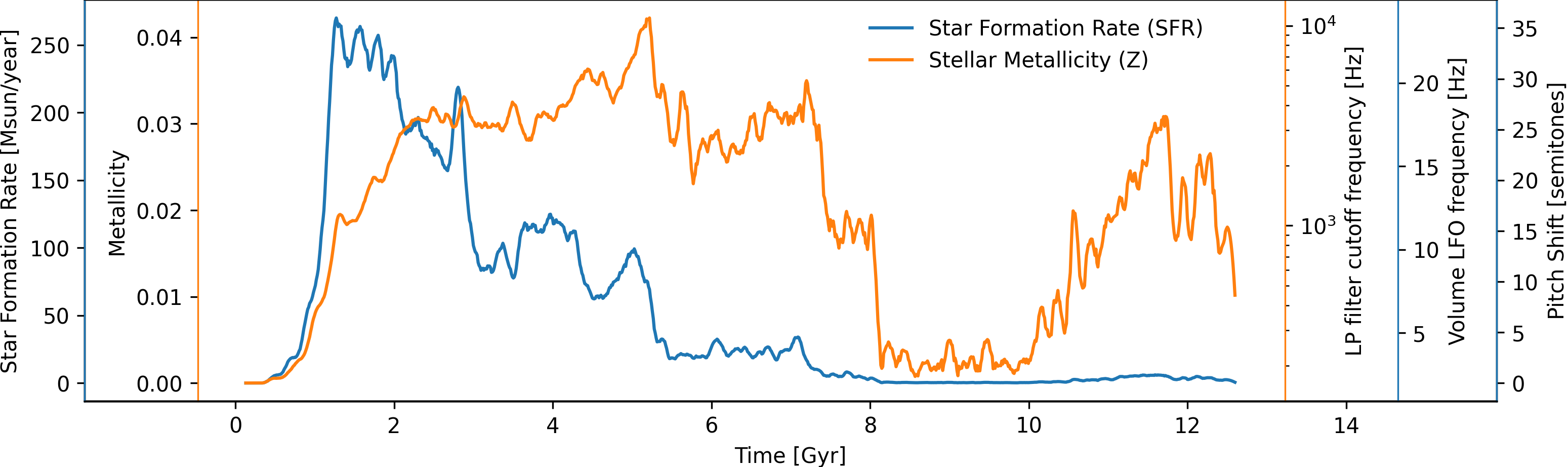}
      \caption{\chg{Example bivariate data series, including: (1) the star formation rate (SFR) and (2) the `metal' mass fraction (known as 'stellar metallicity'), of a simulated galaxy from the \texttt{EAGLE} simulations over 13 Gyr of cosmic time. The metallicity is mapped to a low-pass filter cut off frequency (see right axes), and we provide two different example mappings for simultaneously mapping the star formation rate to another property of the sound. These are pitch and volume LFO frequency, as represented in the axes and described in \S~\ref{sec:multivariate} (where links are also provided to the video files containing both the audio and a visual representation).}}
      \label{fig:multivariate}
\end{figure*}

For demonstrative purposes, we present some example outputs of \strauss{}, showcasing some of the diverse possible applications and some of the key functionality. We include examples of: (1) a univariate data series using 5 different sonification approaches, applicable for research (a galaxy spectrum; \S~\ref{sec:spectrum}); (2) a multiviariate data series using two different approaches, applicable for research (the variation of galaxy properties over time; \S~\ref{sec:multivariate}); and (3) an \chg{aesthetic} spatial sonification application of a series of events for virtual reality (VR), applicable for outreach (the stars appearing in the night sky; \S~\ref{sec:stars}). All application are related to astronomy, but would be applicable to a variety of data with similar formats. While we comment on the utility of these examples, the intention is not to present these as the \textit{`best'} possible representations of the data or the \textit{`proper'} use of \strauss{}. Instead, the idea is to convey the flexibility that \strauss{} can provide, and encourage others to experiment with its features. All the media files associated with the figures and the corresponding sonifications in this work can be found at: \chg{\url{https://data.ncl.ac.uk/articles/media/Trayford_2023_STRAUSS_ICAD_examples/22241182}}, and we just list the relevant filenames below. \chg{a \ttf{README.txt} file also provides details of each file via the aforementioned link.}

\subsection{Research: Flexible Representation of Data Series}
\label{sec:spectrum}

As a demonstration of the diverse possibilities with \strauss{}, we demonstrate multiple audio representations of the single data series plotted in Fig~\ref{fig:spectrum}. This the visible light spectrum of a galaxy taken from the Sloan Digital Sky Survey (SDSS). The data is available from \url{http://sdss.org/}, where the object identification number of this spectrum is: \ttf{1237671260667576428}. These data, show measured \textit{`flux density'} (i.e., energy detected per unit time, per unit area, per unit wavelength) as a function of wavelength. The data exhibit a gradually varying \textit{continuum}, as well as narrow \textit{emission line} peaks, which are the result of particular chemical elements and ions in the galaxy's gas. We may represent this in many ways with \strauss{}. For method (a) we represent each measured data point, within a narrow wavelength range (shaded region in figure) as individual \texttt{Events}. We represent each data point using the packaged \textit{mallets} samples, with an event \textit{time} mapped to wavelength and \textit{pitch} to flux density over a 3 octave range, from an A$_2$ (110~Hz) base note (specified via the \texttt{Score}). The resulting audio file, with an animated figure has been made available in a video mp4 format (\ttf{spectrum\_realisation\_a.mp4}).

An alternative representation can be heard in example (b), for which we can instead think of the spectrum as a single evolving \texttt{Object} (\ttf{spectrum\_realisation\_b.mp4}). Here, we use the \texttt{Synthesizer} type \texttt{Generator}, mapping the pitch shift of an A5 dyad (A$_2$, E$_3$) to flux, smoothly varying with time to represent the varying flux with increasing wavelength. This  uses  the built-in \texttt{pitch\_mapper} preset to configure the \texttt{Synthesizer}. Briefly, this is an unfiltered, single-oscillator triangle wave, with a pitch shift range of 0 to 36 semitones, and all modulation and enveloping turned off. This is an example of using presets to simplify the sonification workflow in \strauss{}. The \texttt{Score} class can be used to specify any choice of notes here - an A5 dyad was chosen as more pleasant than a single note, while avoiding complex harmony that could affect the pitch perception used to convey the data. 

We return to our analogy of common visual representations of data for these sonification approaches; Example (a) is like a \textit{scatter-plot} with individuated data points (see orange points in shaded region of Fig.~\ref{fig:spectrum}), while example (b) is more like a {\em continuous interpolation} of the data points (represented by the blue curve in Fig.~\ref{fig:spectrum}).

We now present examples using the  same continuous, \texttt{Object} representation of (b), but mapping flux density to the cutoff frequency of low-pass filter. This mapping is shown with the right-most axis in Fig.~\ref{fig:spectrum}. This approach effectively varies the \textit{`timbre'} of the sound (here meaning harmonic content) to represent changing flux density, with a \textit{``brighter"} sound corresponding to a \textit{brighter} (i.e., higher flux density) data point.  This allows  us to pick the carrier signal we want without varying its fundamental pitch. For example (c) we use the A5 dyad again, using the default synthesizer preset (3 detuned sawtooth oscillators) for a \textit{tonal} representation with fixed notes that could be used in a harmonious context \ttf{spectrum\_realisation\_c.mp4}). Alternatively, for example (d) we employ a \textit{textural} representation, by varying the filter over a white noise carrier, using the \texttt{windy} built-in preset of \strauss{} (\ttf{spectrum\_realisation\_d.mp4}). We have found that this type of representation can be useful to generate more \textit{listenable} sonification of data series over extended periods \cite{TuckerBrown2022}.

Finally, example (e) is a completely different approach of sound representation (\ttf{spectrum\_realisation\_e.mp4}). Here we map the spectrum shown in Fig.~\ref{fig:spectrum} directly to an audible spectrum, via an inverse Fourier transform. In this regard it may be considered a audification of the data and all the frequency features in the spectrum are heard \textit{``simultaneously"}, with the spectrum represented as one single \texttt{Event}. For this example, we also use the \strauss{} in-built enveloping to apply some \textit{attack} and \textit{decay} to the sonification, for a more organic sound. This affords unique opportunities to explore complex, contemporary datasets with sound, such as hyperspectral data cubes, much more rapidly than is possible by scrolling through independent variable, as above \cite{Trayford23}.

\subsection{Research: Multivariate data series}
\label{sec:multivariate}
\begin{figure*}
\centering
          \includegraphics[width=1.005\linewidth]{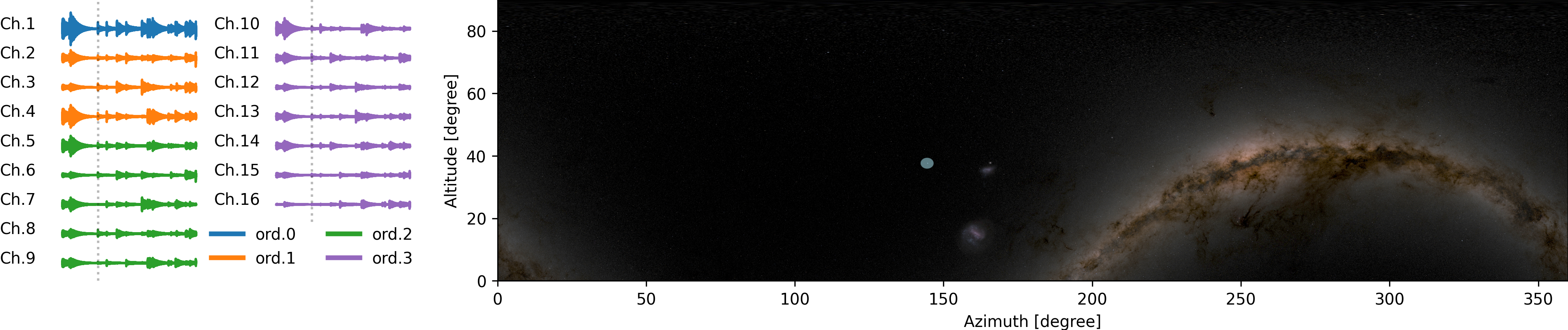}
          \vspace{-4ex}
      \caption{Illustration of an outreach application of \strauss{}, creating spatial audio for the \textit{"stars appearing"} virtual reality exhibit. \textit{Left panel} shows wave-forms of the first 30s of audio in 3rd-order ambisonic format with \textit{ambiX} \cite{Nachbar11} channel ordering. \textit{Right panel} shows a single frame at 8s in equirectangular projection, with a single star appearing (filled circle). Dotted lines mark where this instant corresponds to a triggered sample in the corresponding audio on the left. Described in \S~\ref{sec:stars}, where links are also provided to the video files containing both the audio and a visual representation.}
      \label{fig:stars}
\end{figure*}

The flexible mapping capabilities of \strauss{} allow us to extend the univariate data series  example presented in \S~\ref{sec:spectrum} to \textit{multidimensional} or \textit{multivariate} data. For this example, we use data from the \texttt{EAGLE} simulations of galaxy formation \cite{Schaye15} - using the public \texttt{EAGLE} database \cite{McAlpine16}. EAGLE simulates the formation and evolution of entire populations of galaxies through cosmic time within large, cubic volumes of the Universe and using model prescriptions for unresolved physical processes, reproducing a number of key observation of the real universe.

We selected the galaxy with ID \texttt{21730536} from the largest volume simulation of EAGLE. We concentrate on two separate, but closely linked, data series that describe the star formation history of this galaxy: (1) the rate at which stars formed in the galaxy (SFR); and (2) the `metal' mass fraction of the stars ($Z$) inside the galaxy. The data are visually represented in Fig.~\ref{fig:multivariate}. These two data series provide important insight independently, but how they vary with respect to \textit{each other} also encodes important, complex information about how galaxies evolve over time\footnote{The metals enriching each generation of stars mostly come from former generations dying off as supernov\ae, however these supernova can also drive gas and metals out of galaxies, and fresh gas can fall in from outside the galaxy. The balance of star formation and metal enrichment in a galaxy tells us about this \textit{baryon cycle}.}. Visual inspection of multiple variables at the same time, to gain insight into their relationships, is a ubiquitous problem in physical systems (typically dictated by differential equations). Therefore  auditory inspection of these multiple and evolving properties is a natural avenue for consideration.  

In \strauss{} we can map multiple variables to different expressive properties of sound simultaneously for the same source. In Fig.~\ref{fig:multivariate} we treat the galaxy star formation history as a single \texttt{Object} source as in previous examples, with the \texttt{default} synthesizer preset and using an A5 dyad as carrier, following previous examples. We choose parameter mappings to have some intuitive link to the variables themselves, as indicated in the multiple $y$-axes of Fig.~\ref{fig:multivariate}. We map increasing metallicity $Z$ to the increasing low-pass filter cutoff frequency, yielding a duller sound at low $Z$ and a more `\textit{metallic}' sound at high $Z$ (compare \textit{orange} axes). The star formation rate (SFR) can be thought of as a frequency at which stars form. Therefore, for realisation (a) we map increasing SFR to increasing pitch shift. As before, we produce the audio, along with an animated version of the figure in an mp4 format (\ttf{multivariate\_realisation\_a.mp4}). An alternative realisation (b), maps the SFR to a volume LFO frequency, such that more rapid pulses indicate a higher frequency of star formation (\ttf{multivariate\_realisation\_b.mp4}). \textit{Blue} axes on the right hand side of Fig.~\ref{fig:multivariate} notate these SFR mappings. The efficacy of such multivariate mappings in research contexts requires evaluation, but is naturally possible with \strauss{}. Such mappings can also be used in more impressionistic contexts, discussed next.

\subsection{Outreach: Aesthetic spatial sonification of stars in VR}
\label{sec:stars}

In addition to research application, \strauss{} can be used for more impressionistic sonifications or cases where aesthetics are a higher priority, for example when used for scientific outreach or artistic applications. An outreach application of \strauss{} is exemplified by the \textit{`stars appearing'} VR sequence adapted from the \textit{Audible Universe: Tour of the Solar System} planetarium show \cite{Harrison22}. This example makes use of the native spatial audio output capabilities of \strauss{}, as illustrated in Fig.~\ref{fig:stars}. In this example, $\sim$2500 stars are sonified as \texttt{Events}, mapping their brightness in the observed $V$ band (i.e., using a camera filter which are centred around 550\,nm) to occurrence time, with their base notes mapped to the stars perceived `colour' measured as the difference in brightness of the stars in the $B$ band (a filter centered around 450\,nm) and the $V$ bands. The intention here is to emulate the stars `appearing' to a viewer over the course of a night. We see the brighter stars first, but as our eyes adjust, more and more dim stars appear to us. The colour mapping is also intended to be relatively natural, \textit{`bluer'} stars emitting more \textit{high frequency} radiation are represented by \textit{higher frequency} notes. We use the \texttt{Sampler} generator with Glockenspiel samples, and specify a musical choice of notes via the \texttt{Score}; D$\flat_3$, G$\flat_3$, A$\flat_3$, E$\flat_4$ \& F$_4$ - constituting a D$\flat6/9$ chord. \strauss{} assigns an approximately equal number of stars to each note, D$\flat_3$ is used for stars in the reddest quintile, F$_4$ for the bluest quintile, etc. More details of the data and pitch mapping are provided in \cite{Harrison22}. 

A key feature of this example is the full spatial audio. \strauss{} is able to map the source of the sound accurately to their position on the sky (represented via the altitude and azimuth angles at the observing location) using an ambisonic audio representation. Fig.~\ref{fig:stars} shows the 16 channels representing 3$_{\rm rd}$ order ambisonic audio in \textit{ambiX} format - with the waveforms colour-coded by the ambisonic order they represent. Ambisonics can be naturally represented in \strauss{} as a specific case of the virtual microphone system set up by the \texttt{Channels} class (see \S~\ref{sec:channels}), allowing an arbitrary number of sources to be represented as part of an accurately spatialised soundscape. Inspecting the waveforms in Fig.~\ref{fig:stars} shows how the audio triggered by a single star appearing contributes differently to each of the ambisonic channels. This example can be experienced in a variety of formats, including: (1) the full 3rd order audio included as a \texttt{.wav} file (\ttf{rendered\_stars\_ambix3.wav}); (2) a file mixed down to 8 channels for standard VR playback support (e.g. via \textit{Oculus TV}, \textit{DeoVR}, etc; \ttf{stars\_appearing\_mw\_ambiX3\_360\_FB360.mkv}); and (3) a video that shows the first person view (`PoV'; though with sound mixed down to mono; \ttf{stars.mp4}).

\section{Summary \& Future Work}
\label{sec:summary}

We present the motivation and philosophy for the sonification Python package \strauss{}, outlining the general structure of the code and demonstrating some example applications that showcase the diverse  features in action. This is intended to be a broad introduction to \strauss{} and its key concepts, acknowledging that the code is a work in progress, with plans for significant development. To complement this, the online \strauss{} repository and documentation serve as dynamic documents, providing up-to-date technical details on the operation of the code, relevant examples, and listing planned features. Going forward, we are beginning a period of dedicated development of the code, which will allow many of the feature presented here to be developed further and many new features to be added, along with user interface and accessibility enhancements.

Some of the focus points for future development relate to the \textit{"low barrier, high ceiling"} philosophy and accessibility features planned for \strauss{}. In particular, the \textit{presetting} (or \textit{preconfiguration}) approach  demonstrated in the \ttf{Generator} and \ttf{Channels} classes, could be extended to the remaining classes in the code, allowing users to choose from representative lists of (editable) configurations to try for their data, as a beginners' entry-point to using \strauss{}. We also plan extensions for real-time audio rendering, and to make use of existing digital sound resources. Generally, \strauss{} development will be responsive to the features and improvements that users desire, welcoming community contributions in the free, open source (FOSS) paradigm. We hope that making \strauss{} available to the astronomy community, and data-centric fields more broadly, will lead innovative code applications and sonifications to naturally emerge. 

\section{Acknowledgements}
\label{sec:ack}
JWT acknowledges support via the \textit{STFC Early Stage Research \& Development Grant}, reference ST/X004651/1.
CMH acknowledges funding from an United Kingdom Research and Innovation grant  (code: MR/V022830/1).

\bibliographystyle{IEEEtran}
\bibliography{refs2023}

\end{sloppy}
\end{document}